\documentclass[11pt]{article}
\usepackage{fullpage}
\usepackage{float}
\usepackage{graphicx}
\usepackage{amsmath}
\usepackage{subcaption}
\usepackage{bm}
\usepackage[dvipsnames]{xcolor}
\usepackage[colorlinks=true,linkcolor=blue, citecolor = blue]{hyperref}
\usepackage[round]{natbib}

\providecommand{\keywords}[1]
{	
  \textbf{Keywords ---} #1
}

\title{Network Analysis of Orchestral Concert Programming}
\author{Anna K. Yanchenko\\ Department of Statistical Science \\ Duke University}
\date{}

\begin{document}
\maketitle


\begin{abstract}
Orchestral concert programming is a challenging, yet critical task for expanding audience engagement and is usually driven by qualitative heuristics and common musical practices.  Quantitative analysis of orchestral programming has been limited, but has become more possible as many orchestras archive their performance history online.  The contribution of this work is to use statistical network models to quantitatively explore orchestral concert programming, focusing on which factors determine if two composers are programmed together in the same concert by the Boston Symphony Orchestra. We find that the type of composition is the most important covariate in determining which composers are performed together and the additive and multiplicative effects are logical from an orchestral programming perspective.  These results suggest that a network analysis is a promising approach for the analysis of concert programming, with several directions for future extensions.
\end{abstract}


\section{Introduction}\label{sec:introduction}

Orchestral concert programming involves selecting pieces to be performed, both for specific concerts and across an entire season.  Concert programming is one of the most important and challenging jobs for a symphony's artistic or music director.  Programs must be coherent, both for a single concert and across an entire season, without the repetition of too many pieces from one season to the next.  Music directors must consider both the musical components of a concert, balancing the tone, style and lengths of the pieces performed, as well as more financial demands, such as the cost of the personnel required for the pieces performed.  In addition to appealing to the musical tastes of the orchestra's subscribers, music directors often look to push the boundaries of audience expectations through the performance of new works or rarely-performed pieces \citep{Wittry:2007}.  However, music directors must also look to expand audience engagement beyond current subscribers, especially with the recent decline in orchestral attendance \citep{Midgette:2005}.  ``Pops'' concerts, thematic programs and famous soloists are common ways that music directors seek to grow orchestral audiences.\newline

\noindent While orchestras look to improve audience engagement for future and current subscribers, there is also the desire to balance new works with the standard repertoire.  There has been criticism of the overly ``conservative'' programming of most large orchestras \citep{Tommasini:2008}.  Concert programs tend to follow the traditional format of an ``overture, concerto and symphony'' per concert and there is often little experimentation with non-traditional pairings of pieces or the performance of new works.  Additionally, most new works are not performed again after their premiere.  Especially as audience attendance at classical music concerts decreases, the importance of orchestral programming increases.\newline

\noindent Several established guidelines and suggestions for concert programming exist from a predominantly musical perspective, for example, \citet{Goza:2006} and \citet{Wittry:2007}.  Additionally,  \citet{Gilmore:1993} and \citet{Thuerauf:2005}  studied concert programming  across major orchestras for the 1969 - 1970 and 2003 - 2004 seasons, respectively, and both found that there were a limited number of composers (i.e. Mozart, Beethoven and Tchaikovsky) and a limited number of musical eras (i.e. Romantic era and 20th Century music) that dominated orchestral concert programming.  The author of \citet{Gilmore:1993} in particular used these findings to advocate for the performance of more new works and a shift away from the standard, established repertoire of pieces. Both \citet{Gilmore:1993} and \citet{Thuerauf:2005} provided summary statistics and performed analysis of variance (ANOVA) tests for concert programming, though prior quantitative analysis of concert programming is sparse.\newline

\noindent This work is a preliminary analysis that uses statistical network models to quantitatively explore orchestral concert programming, focusing on which factors determine if two composers are programmed together in the same concert.  We use data from the Boston Symphony Orchestra from the 1999-2000 to 2017-2018 seasons, which includes 2464 unique concerts and 330 distinct composers.  To our knowledge, this is one of the first model-based, network analyses of orchestral concert programming data in general, and one of the first quantitative explorations of the Boston Symphony Orchestra concert archives, specifically.  We find that statistical network models are well suited to modeling orchestral programming data and that the models recover several widely-held views of orchestral programming.


%
\section{Data and Exploratory Data Analysis}\label{sec:EDA}


\subsection{Boston Symphony Orchestra}

The orchestral concert programming data consists of over 2464 unique concerts performed by the Boston Symphony Orchestra (BSO) from the 1999-2000 season through the 2017-2018 season \citep{BSO}.  The data comes from the BSO concert archives, which has information about every concert performed in the orchestra's history \citep{BSO}.  The analysis for this work focuses on composer-level traits, rather than on individual pieces.  Since there are only a few pieces that are performed every year or even every other year, the data is too sparse to model at the piece-level.  However, there are many composers that are performed every year, allowing for enough density of data at the composer-level. \newline

\noindent There are 330 unique composers considered in the data.  For this work, the composer-level covariates considered are the year of birth of the composer (which serves as a numeric proxy for the era of the composer), the nationality of the composer by region and the type of piece performed by each composer (overture, concerto, symphony, or other).

\subsection{Exploratory Data Analysis for the BSO}

\begin{figure}
 \centerline{
 \includegraphics[width=0.8\columnwidth]{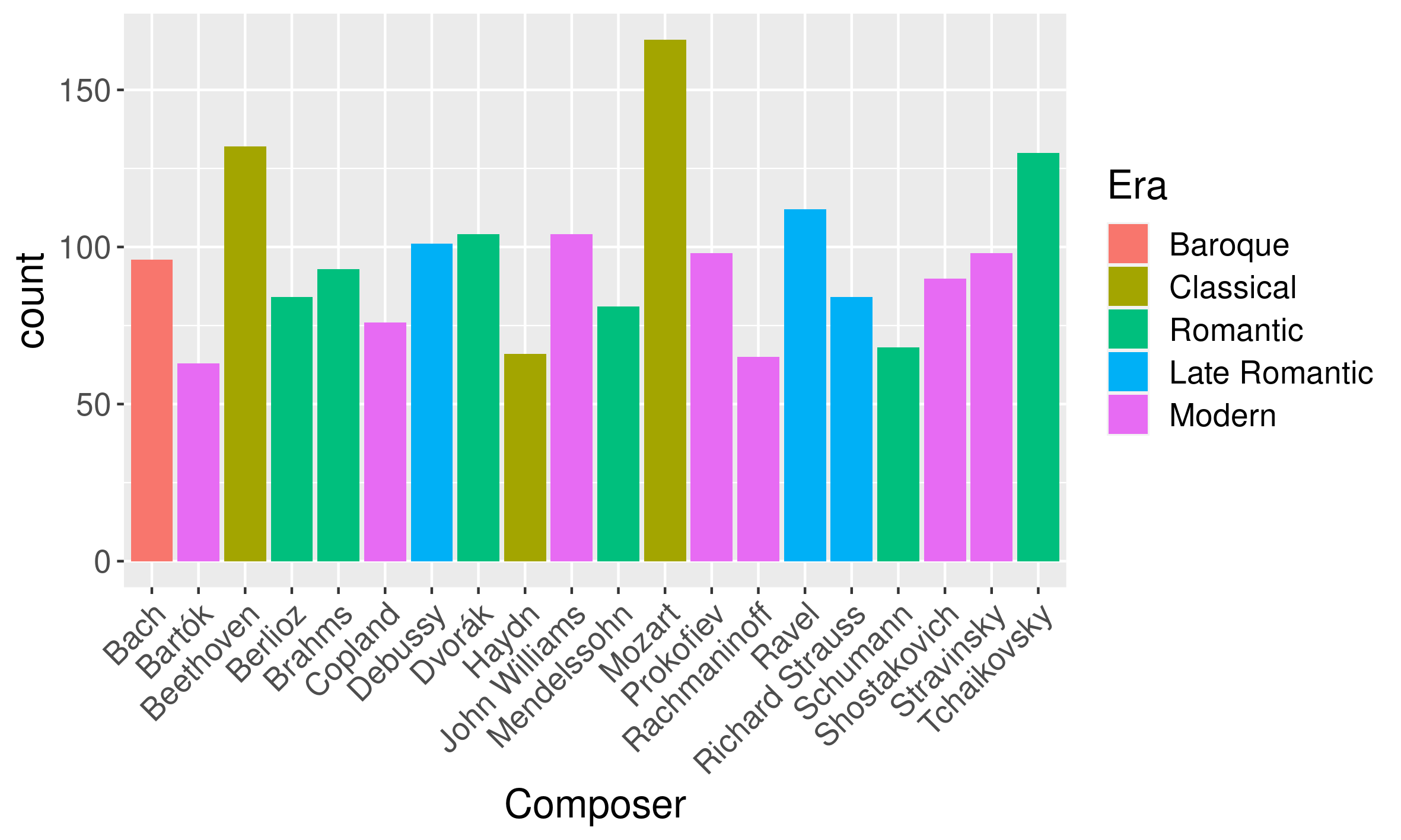}}
 \caption{Top 20 most frequently performed composers, colored by musical era.}
 \label{fig:comps}
\end{figure}

An exploration of the data confirms several conventional wisdoms about the types of composers that were programmed by the BSO since the 1999-2000 season.  Overall, the composers performed most often included the most popular Western composers, such as Mozart, Bach and Beethoven (\autoref{fig:comps}).  Interestingly, John Williams was also one of the most frequently performed conductors, likely due to his popularity in Pops style concerts and the fact that he is the conductor emeritus of the sister Boston Pops Orchestra.  Shostakovich was also performed more frequently than may be expected for a ``typical'' orchestra.  The BSO is currently recording all of the Shostakovich symphonies under music director Andris Nelsons, and thus the composer has been performed frequently, especially recently.\newline

\begin{figure}
\centering
\begin{subfigure}{.45\textwidth}
  \centering
  \includegraphics[width=\linewidth]{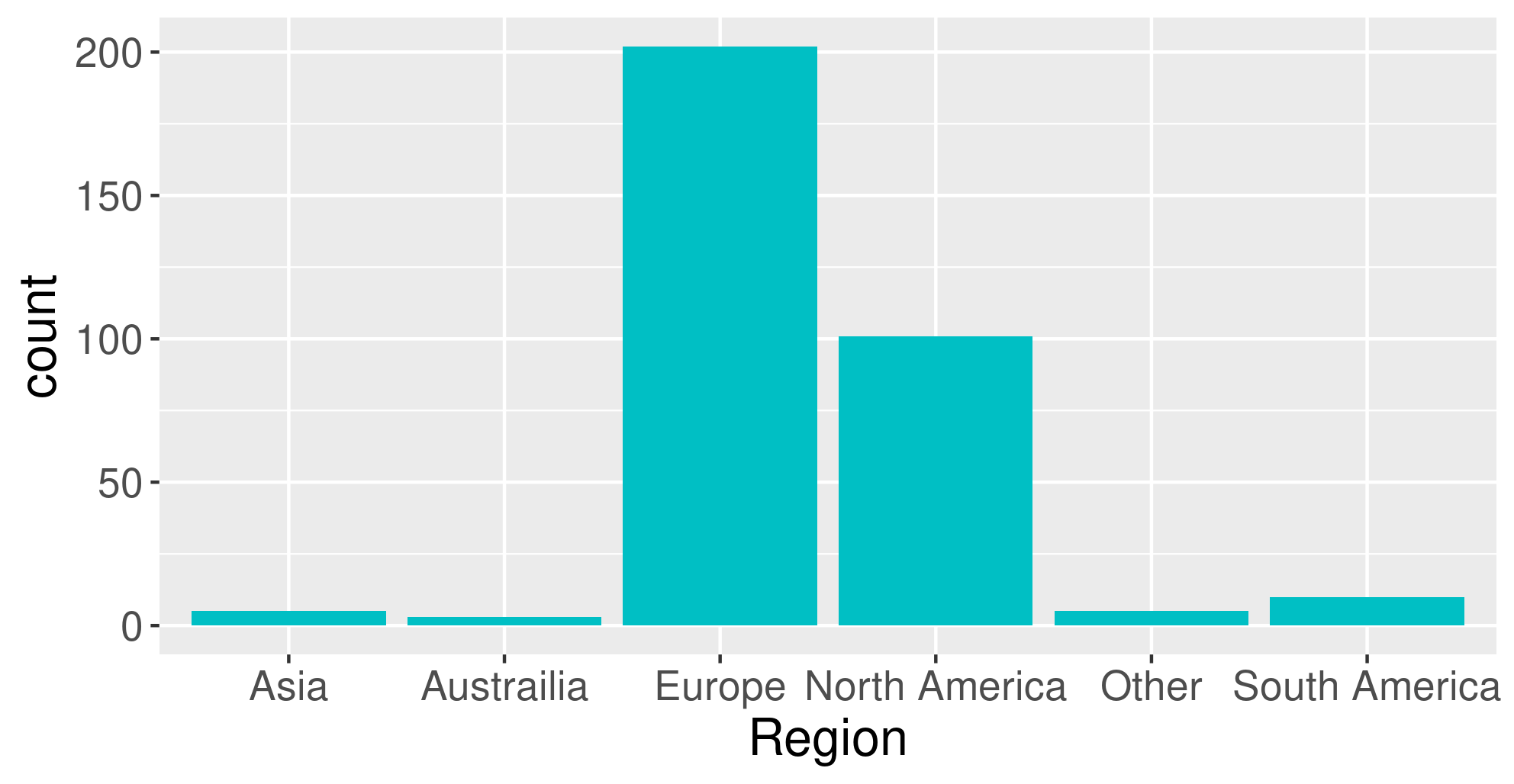}
  \caption{Nationality.}
  \label{fig:region}
\end{subfigure}%
\begin{subfigure}{.45\textwidth}
  \centering
  \includegraphics[width=\linewidth]{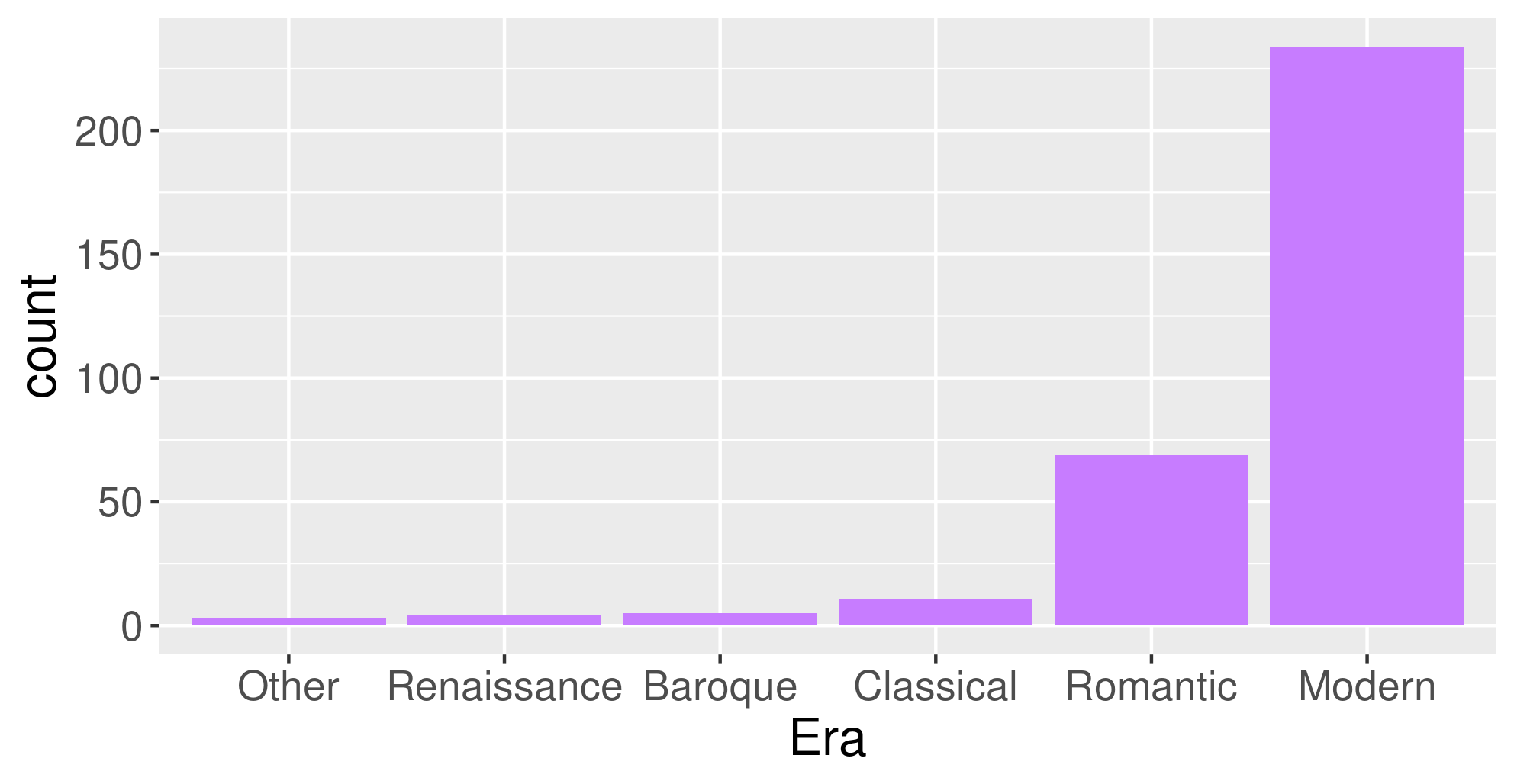}
  \caption{Era.}
  \label{fig:era}
\end{subfigure}
\caption{(a) Nationality by region and (b) musical era of composers performed by the BSO.}
\label{fig:chroma}
\end{figure}

\begin{figure}
 \centerline{
 \includegraphics[width=0.7\columnwidth]{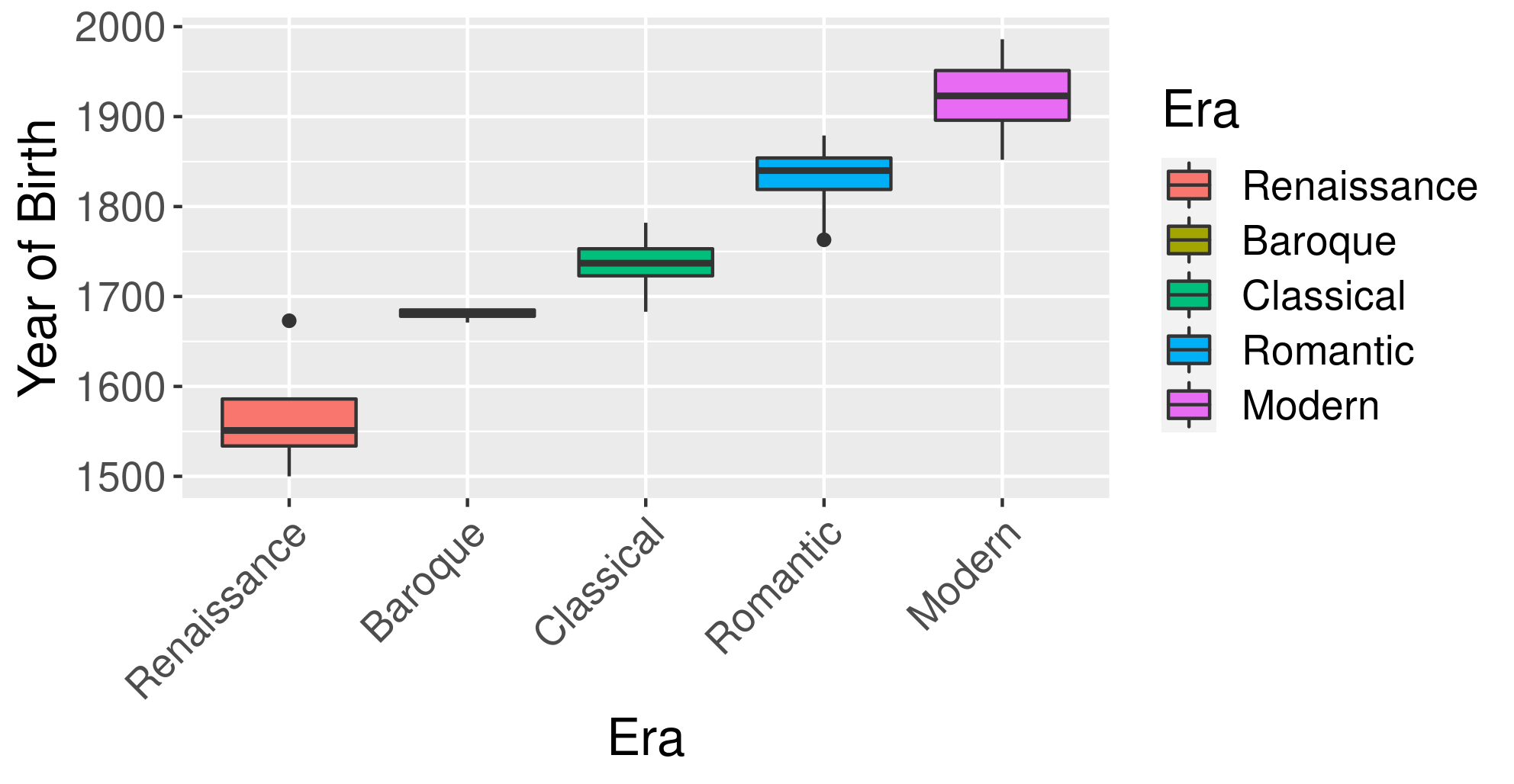}}
 \caption{Musical era by year of birth of composers performed by the BSO.}
 \label{fig:era-DOB}
\end{figure}

\noindent The majority of composers performed were European, though there were also many North American composers performed (\autoref{fig:region}).  Romantic and Modern were the most common musical eras of pieces performed (\autoref{fig:era}).  However, as musical era can be subjective for some composers, the year of birth of each composer was included as a covariate instead, as year of birth is highly correlated with the musical era of the composer (\autoref{fig:era-DOB}).  The majority of composers performed were born after 1800.  For both the nationality by region and the musical era of the composer, the category ``Other'' refers to traditional or anonymous pieces, such as \textit{Shenandoah}, that do not have a specific, attributed composer.  Such pieces were grouped into an ``Anonymous'' composer category, and assigned a default year of birth.  \newline

\noindent Finally, as orchestral concerts have a finite length, the type of piece performed is a very important covariate in determining which composers and pieces will be performed together in a specific concert.  The piece is grouped into one of four categories: ``Overture'', ``Concerto'', ``Symphony'' or ``Other'', based on the name of the piece.  That is, if the piece has ``Overture'' in the name, it is grouped into the ``Overture'' category.  All pieces that do not have either ``Overture'', ``Concerto'' or ``Symphony'' in the name of the piece are grouped into the ``Other'' category.  This is a simplifying assumption, as there are many pieces, especially modern pieces, that do not include the type of piece in the title.  However, this approach results in an objective grouping by type of piece and future work will look at a more nuanced categorization of type of piece, perhaps based on the length of the piece.  The largest category of type of piece performed was ``Other'' by a large margin (partially due to the nature of the grouping), which accounted for 62\% of the pieces performed.  The second largest category was ``Concerto'' (18\%), followed by ``Symphony'' (12\%) and ``Overture'' (8\%).  Code and the data after pre-processing is available at \url{https://github.com/aky4wn/Network-Programming}.



\section{Network Model}\label{sec:model}

The goal of this work is to explore which composers are programmed together in a specific concert, which requires capturing the relationships between different composers programmed in the same concert.  Network models provide a natural model choice to explicitly model the relationships between composers and to give insight into which qualities of composers make them more likely to be programmed together in the same concert.  In this section, we give an overview of statistical network models, describe the pre-processing necessary to transform the BSO data into a network structure and finally describe the specific network model used to analyze the BSO concert data.

\subsection{Network Modelling}

Network models explicitly capture quantitative relationships between different entities of interest.  Each entity is called a node in the network and different nodes are connected via edges, which describe the relationship between the entities.  These edges can be directed or undirected, depending on the nature of the relationship between nodes.  Additionally, each node can have covariates that describe various node-specific quantities of interest and the edges between nodes that describe the relationship can be binary, categorical or continuous. Data with a network structure requires specific assumptions about the error structure of a given statistical model.  For example, nodes are not independent of each other, so the error in modeling the relationship between node $i$ and node $j$ is correlated with the error in modeling the relationship between node $j$ and node $i$ \citep{Hoff:2018}.  This type of dependency between nodes cannot be accurately captured by generalized linear models, for example.  Network models have been widely used in a variety of different fields, in particular, political science and social networks analysis \citep{Hoff:2018}.   Network models have been previously used to explore general relationships between classical composers in \citet{Park:2015}.

\subsection{Data Pre-Processing}

\begin{figure}
 \centerline{
 \includegraphics[width=0.35\columnwidth]{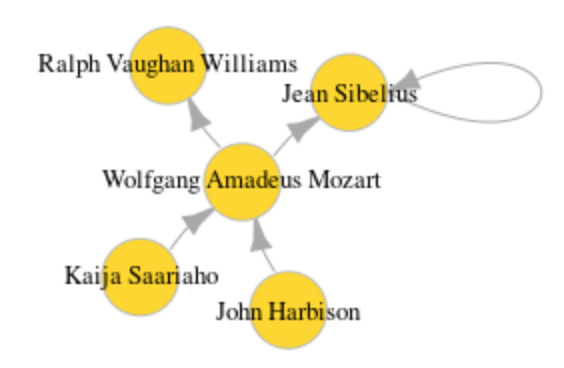}}
 \caption{Example directed network for the BSO programming data. The arrows indicate the order in which a composer was performed. For example, Mozart was performed after Saariaho and Harbison and before Vaughan Williams and Sibelius.}
 \label{fig:network}
\end{figure}

For the BSO concert data, each composer is a node in the network and the outcome variable of interest is a binary random variable, $Y_{ij}$, where $Y_{ij} = 1$ if composer $i$ was performed before composer $j$ in any concert in the time span considered.  This results in a directed network, an example of which is shown in \autoref{fig:network} for a small subset of composers.  The order in which composers are performed in a concert is important and modeling a directed network allows for a concert where composer $i$ was performed before composer $j$ to be distinct from a concert in which composer $j$ was performed before composer $i$.  We consider this assumption from a modeling perspective below. \newline

\noindent As described in Section \ref{sec:EDA} above, we include the following nodal, composer-level covariates: year of birth of the composer, nationality by region and type of composition performed.  The year of birth is a continuous variable and the nationality by region is categorical.  Each type of piece, ``Overture'', ``Concerto'', ``Symphony'' and ``Other'', is a binary variable that is aggregated over all performances in the time period considered for that specific composer.  For example, for the composer Tchaikovsky, the nodal covariates are as follows: Nationality by Region = European, Year of Birth = 1840, Overture = 1, Concerto = 1, Symphony = 1, Other = 1.  The BSO performed at least one overture, concerto, symphony and other (for example, the Suite from the Nutcracker) type of piece by Tchaikovsky  between the 1999-2000 season and the 2017-2018 season. Again, this is a simplifying assumption and future work will look at including the frequency of performance more explicitly in the model.\newline

\noindent Unlike the exploratory analysis conducted in Section \ref{sec:EDA} above, this network structure of the concert programming data gives additional information about the relationship between composers. For example, we can analyze the in-degree and out-degree of each composer performed.  The degree of a node is the number of distinct edges related to that node.  For a given node $i$, the in-degree indicates how many distinct composers were performed before composer $i$ and the out-degree describes how many distinct composers were performed after composer $i$.  The top 10 composers with the largest in- and out-degrees is given in \autoref{tab:EDA}.  Mozart and Beethoven were performed both before and after a large number of distinct composers, indicating that Mozart and Beethoven were both performed frequently (as confirmed in Section \ref{sec:EDA}), but also that they were programmed with a wide variety of different composers in the same concert.  A composer like Stravinsky, on the other hand, was programmed before less than half the number of distinct composers as Mozart, reflecting that many pieces by Mozart have more flexibility for the other types of pieces and composers that they can be programmed with, as compared to Stravinsky.  This relational information is important to understanding how a program is put together and can be reflected and captured by a network structure model, as opposed to just summary statistics of the data. 

\begin{table}
 \begin{center}
 \begin{tabular}{|l|l||l|l|}
  \hline
 Composer & Out-Degree & Composer & In-Degree \\
  \hline\hline
  Mozart & 59 & Mozart & 57 \\\hline
  Beethoven & 47 & Beethoven & 56 \\\hline
  Ravel & 38 &  Tchaikovsky & 49 \\\hline
  Haydn & 34 & Dvor\'ak & 39 \\\hline
  Bach & 32 & Stravinsky & 38 \\\hline
  Brahms & 32 & Prokofiev & 38 \\\hline
  Tchaikovsky & 31 & Ravel & 37 \\\hline
  Debussy & 30 & Brahms & 35 \\\hline
  Prokofiev & 28 & Shostakovich & 31 \\\hline
  Stravinsky & 27 & Mahler & 31 \\\hline
 \end{tabular}
\end{center}
 \caption{Top 10 composers with the largest out- and in-degrees in the BSO network. These composers were performed before (out-degree) or after (in-degree) the largest number of other composers.  For example, Mozart was performed before 59 other distinct composers and after 57 distinct composers.}
 \label{tab:EDA}
\end{table}

\subsection{AME Model}

We use a probit social relations regression model with additive and multiplicative effects (AME model) \citep{Hoff:2018, Hoff:2015} for the BSO programming data.  This model is selected for its ability to capture higher-order dependencies between composer nodes \citep{Hoff:2018}. The model likelihood is given in Equation \ref{eq1}.  The binary outcome variable of interest is $Y_{ij}$, where $Y_{ij} = 1$ if composer $i$ was performed before composer $j$.  Let $n$ be the number of distinct composers. The network data can be expressed as an $n$ x $n$ adjacency matrix, where the entry in row $i$ and column $j$ corresponds to $Y_{ij}$.  The probit regression model specifies $z_{ij}$ as a latent variable that determines if $Y_{ij}$ is equal to 1 or 0 through the indicator function.  The nodal covariates are the nationality by region, year of birth and type of piece for each composer.  The nodal covariates for the rows of the adjacency matrix are $X_r$ and the nodal covariates for the columns of the adjacency matrix are $X_c$, where $X_r = X_c$ since the same composers are represented on the rows and columns of the adjacency matrix.
\begin{equation}\label{eq1}
  \begin{split}
    z_{ij} &= \beta_r^TX_{ri} + \beta_c^TX_{cj} + a_i + b_j + u_i^Tv_j + \epsilon_{ij}\\
    Y_{ij} &= \bm{1}(z_{ij} > 0)\\
  \end{split}
\end{equation}

\noindent The row and column regression coefficients, $\beta_r$  and $\beta_c$, respectively, describe the relationship between the nodal covariates and whether composer $i$ was programmed before composer $j$. The additive row,   
$\{a_i\}_{i=1}^n$, and column, $\{b_j\}_{j=1}^n$, effects capture the number of other distinct composers programmed after composer $i$ and before composer $j$.  For example, a large value for $b_j$ indicates that composer $j$ was frequently performed after a large number of distinct composers, while a small value $b_j$ indicates that composer $j$ was only performed after a very small subset of other composers.  The $\{u_i\}_{i=1}^n$ and $\{v_j\}_{j=1}^n$ are the row and column multiplicative effects, respectively.  The multiplicative effects can be viewed as latent factors that allow the model to capture higher order network dependencies between composers, for example, triad dependencies (relationships between 3 distinct composers) rather than just dyad dependencies (relationships between only 2 composers) \citep{Hoff:2018}.  The dimension of the latent space is $R = 3$.

\begin{equation}\label{eq2}
\begin{split}
(a_i, b_i) &\overset{iid}{\sim} \mathcal{N}(0, \Sigma_{ab}), \; i = 1, \ldots, n \\
\{(\epsilon_{ij}, \epsilon_{ji}) : i\neq j\} &\overset{iid}{\sim} \mathcal{N}(0, \Sigma_{\epsilon}) \\
(u_i, v_i) &\overset{iid}{\sim} \mathcal{N}(0, \Psi), \; i = 1, \ldots, n \\
\Sigma_{ab} &= \begin{bmatrix}
\sigma^2_a & \sigma_{ab}\\
\sigma_{ab} & \sigma^2_b
\end{bmatrix} \\
\Sigma_{\epsilon} &= \sigma^2_{\epsilon}\begin{bmatrix}
1 & \rho\\
\rho & 1
\end{bmatrix} \\
\end{split}
\end{equation}

\noindent We perform Bayesian inference for parameter estimation \citep{BDA}.  Following \citep{Hoff:2018}, the priors for the parameters of interest are specified in Equation \ref{eq2}. The probit model assumes Gaussian errors for $\epsilon_{ij}$; $\mathcal{N}(\mu, \Sigma)$ represents the normal distribution with mean $\mu$ and covariance $\Sigma$.  The variance structure of the AME model includes $\sigma_a^2$ and $\sigma_b^2$, which represent across-row and across-column heterogeneity, respectively, $\sigma_{ab}$, which represents the linear association between row and column means, $\sigma_{\epsilon}^2$, which represents additional variability across dyads (pairs of composer nodes that are connected) and $\rho$, which is the within dyad correlation that is not explained by $\sigma_{ab}$ \citep{Hoff:2018}.    Following \citep{DIY:2018, AMEN:2017}, standard hyper-priors are used for $\beta_r$, $\beta_c$, $\rho$ and $\sigma_{\epsilon}^2$ and empirical Bayes estimates from the data are used to specify the hyper-priors for $\Sigma_{ab}$ and $\Psi$. The model was fit in R using the AMEN package \citep{AMEN:2017}.

\subsection{Model Fit}

Two forms of network structure are considered to explore and compare various model fits. The first is a symmetric graph structure, where the order in which composers are performed in the same concert does not matter.  The second is an asymmetric graph structure where the order of performance does matter. That is, $Y_{ij} = 1$ if composer $i$ is performed before composer $j$ in the same concert; this is the final model considered in the analysis and is described in the preceding section.  Additionally, several variations of the full model described above are considered: the Simple Random Graph (SRG) model, $z_{ij} = \mu + \epsilon_{ij}$, which only contains an intercept term, the Social Relations Model (SRM),  $z_{ij} = \mu + a_i + b_j + \epsilon_{ij}$, which only contains additive random effects and no covariates, the Social Relations Regression Model (SRRM), $z_{ij} = \beta^T_rX_{ri} + \beta^T_cX_{cj} + a_i + b_j +  \epsilon_{ij}$, with additive effects and covariates, a logistic regression model, $z_{ij} = \beta^T_rX_{ri} + \beta^T_cX_{cj} + a_i + b_j +  \epsilon_{ij} $, which assumes the error terms are iid and thus does not take into account the network structure of the data, and finally the full Additive and Multiplicative Effects (AME) model, \autoref{eq1}.  The dimension of the multiplicative effect vectors is $R = 3$. All models are fit using the AMEN package in R \citep{AMEN:2017}.\newline

\begin{figure}
 \centerline{
 \includegraphics[width=0.8\columnwidth]{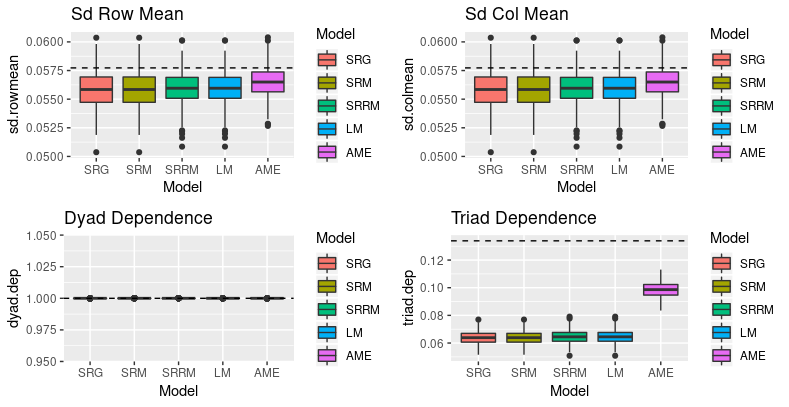}}
 \caption{Goodness of fit checks for the symmetric graph models. The horizontal black lines represent the observed value of each statistic, while the boxplots for each model are posterior predictive samples. }
 \label{GOF-sym}
\end{figure}

\begin{figure}
 \centerline{
 \includegraphics[width=0.85\columnwidth]{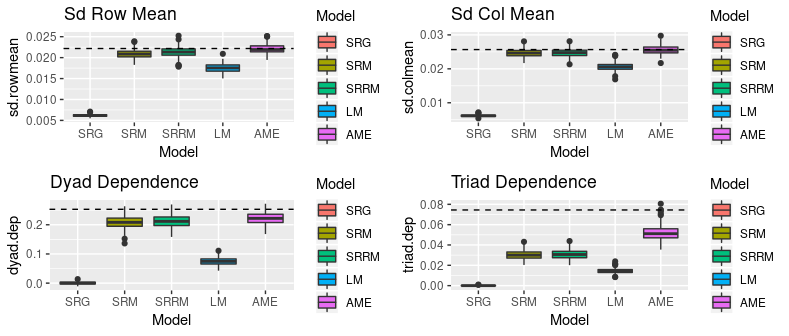}}
 \caption{Goodness of fit checks for the asymmetric graph models. The horizontal black lines represent the observed value of each statistic, while the boxplots for each model are posterior predictive samples. The asymmetric graph models better capture higher order dependencies in the network structure, compared to the symmetric models in \autoref{GOF-sym}.}
 \label{GOF-asym}
\end{figure}

\noindent Observed values and posterior predictive draws for four goodness of fit statistics are given in \autoref{GOF-sym} for the symmetric graph and  \autoref{GOF-asym} for the asymmetric graph.  The goodness of fit statistics are the standard deviations of the row and column means of the network, the dyad dependence and the triad dependence.  The dyad dependence is the correlation between the rows and columns of the adjacency matrix for the network, and so is identically 1 for the symmetric network. The AME model achieves the best fit for all four statistics for the asymmetric network model.  The triad dependence is still not extremely well captured by this model, but the posterior predictive draws are closest to the observed value.\newline 

\noindent As expected, the baseline SRG and logistic regression models do not model either the symmetric or asymmetric networks very well and are not able to capture dyad or triad dependencies in either network, as the model structure does not consider this more complex data structure.  
The AME model achieves the best model fit in terms of these statistics on the asymmetric network. Thus, it is important to include composer nodal covariates and additive and multiplicative random effects. The order in which composers are performed is also important, as this asymmetric structure is better captured by the AME models considered than the symmetric network structure.  All remaining results will be presented in the context of the AME model on the asymmetric network.

\section{Results}\label{sec:results}

In this section, we analyze the results of the AME model on the BSO concert programming data, focusing on the regression coefficients, additive random effects and multiplicative random effects.  We find that the AME model is able to recover several expected results in terms of conventional wisdom for orchestral concert programming, as well as suggest some directions for further analysis.\newline

\begin{figure}
 \centerline{
 \includegraphics[width=0.7\columnwidth]{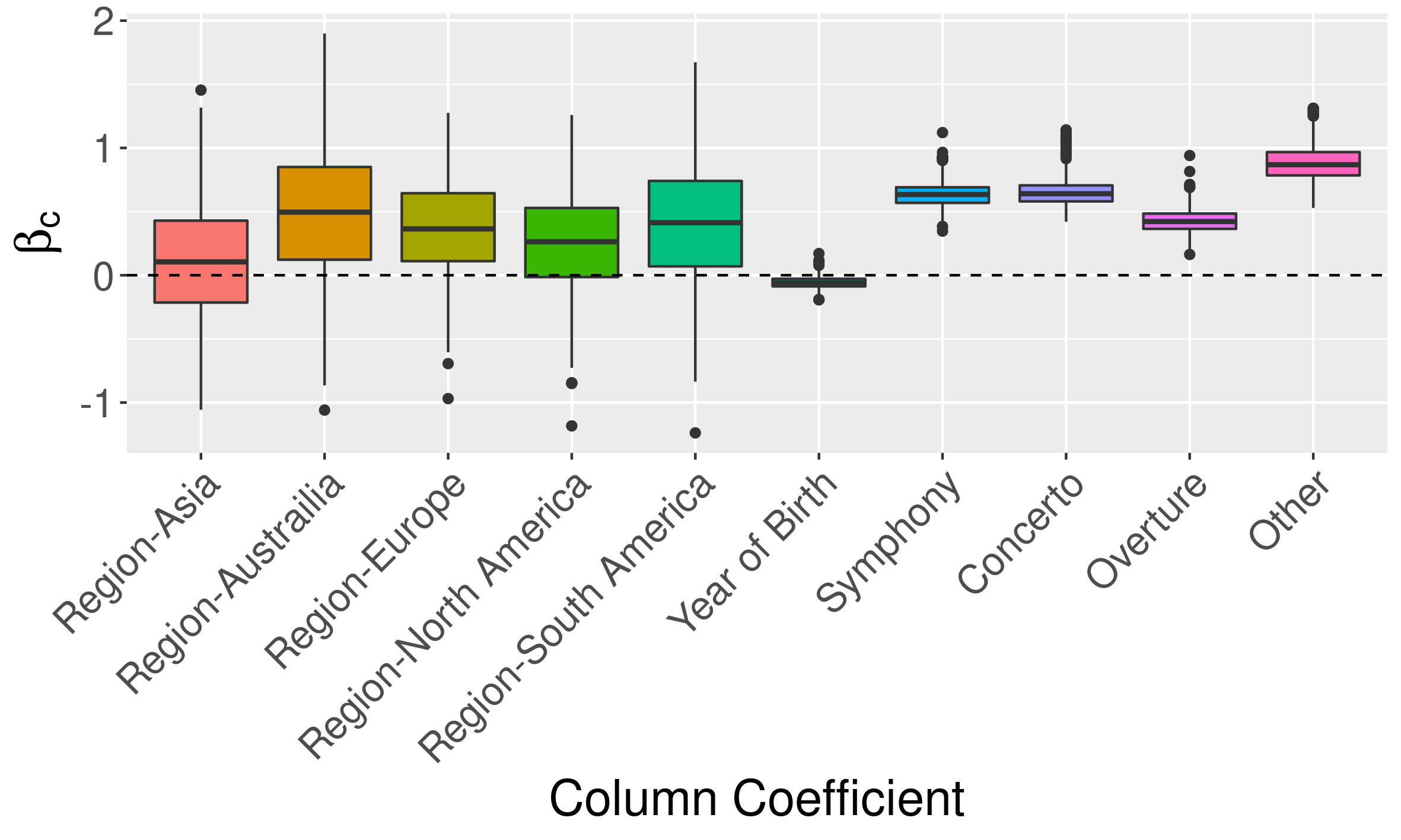}}
 \caption{Posterior distributions for the column regression coefficients, $\beta_c$.  The only significant covariates (i.e. those whose posterior distributions do not contain 0) are the type of piece covariates.  The p-values for the ``Symphony'', ``Concerto'', ``Overture'' and ``Other'' coefficients were all 0.  All other coefficients had p-values larger than 0.2.}
 \label{fig:beta}
\end{figure}

\noindent The regression coefficients, $\beta_r$ and $\beta_c$ in Equation \ref{eq1} indicate how much each nodal, composer-level covariate impacts whether two composers were programmed together.  The posterior distributions for the $\beta_c$ regression coefficients are shown in \autoref{fig:beta}; the posterior distributions for the $\beta_r$ coefficients are nearly identical.  The posterior distributions for the nationality by region and the year of birth of the composers contain 0, and are thus not significant in determining if two composers were programmed together (the ``Other'' Nationality region is the reference category).  This is somewhat surprising, as the year of birth is a proxy for the era of the composer, and there are many concerts that have pieces and composers all from the same musical era.  Based on the AME model fit, the type of piece coefficients are the most significant for all four types of pieces.  This result makes sense, as a program that contains four symphonies is very unlikely, due to the finite length of concerts.  Thus, the importance of the type of piece in determining if two composers are programmed together is confounded by the length of each piece.  However, overall the type of piece is the most important covariate of those considered in terms of determining if two composers will be programmed together in the same concert.\newline

\begin{figure}
 \centerline{
 \includegraphics[width=0.7\columnwidth]{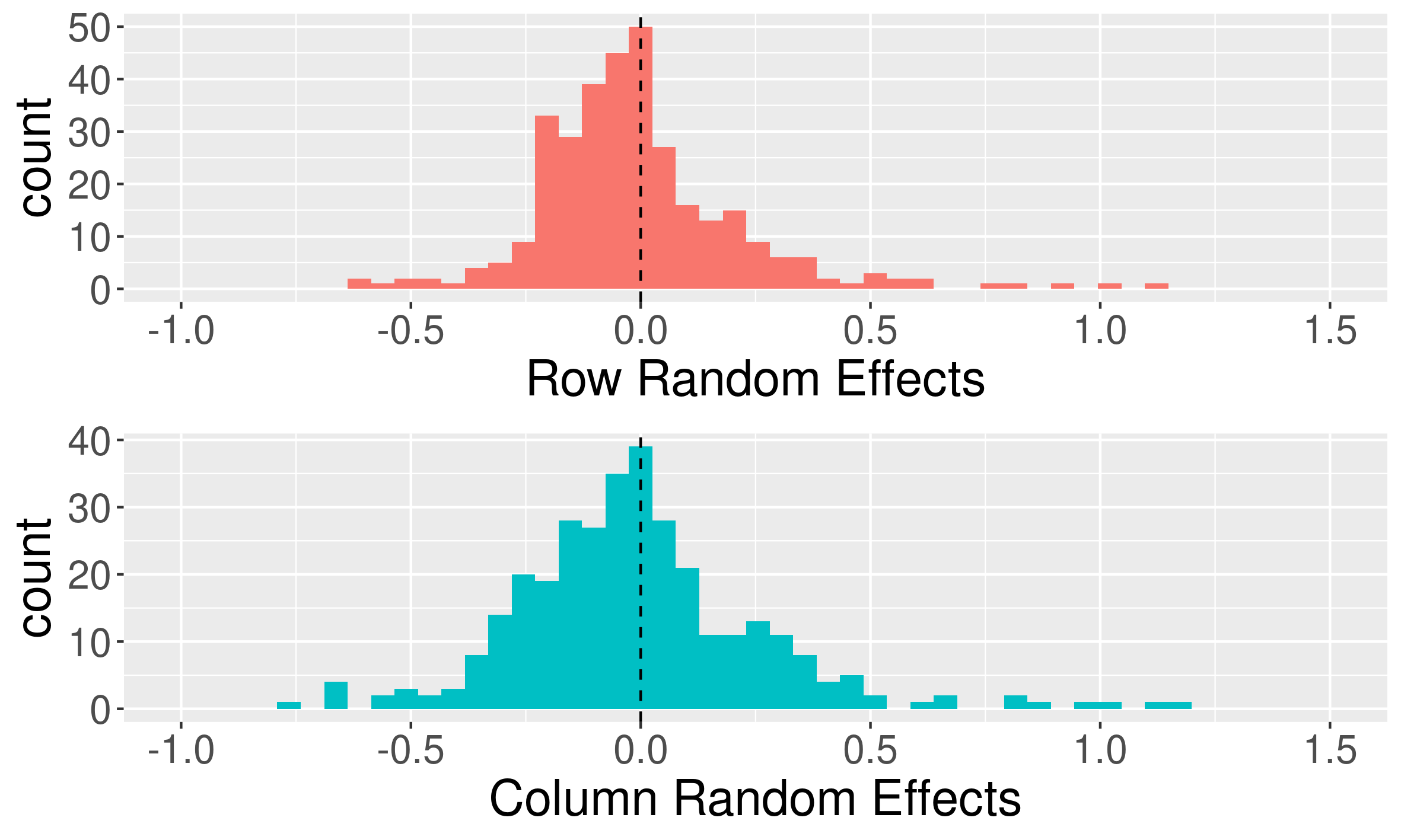}}
 \caption{Posterior means of the row ($a_i$) and column ($b_j$) additive random effects, across all composers.}
 \label{fig:add}
\end{figure}

\noindent The majority of the additive row ($a_i$) and column ($b_j$) random effects are relatively small and centered about 0 (\autoref{fig:add}).  However, there are some composers that do have large posterior mean additive random effects (\autoref{tab:row-effects-large}).  For example, the composers Debussy, Ravel and Bach have the three largest posterior mean row random effects, indicating that these composers were frequently performed and were performed before many other composers.  It is somewhat surprising that these 10 composers are programmed before the largest number of other composers, however, many of these composers have shorter pieces that can be performed prior to a variety of composers.\newline  

\begin{table}
 \begin{center}
 \begin{tabular}{|l|l|}
  \hline
 Composer  &  $a_i$ \\
  \hline\hline
  Claude Debussy & 1.51 \\\hline
  Maurice Ravel & 1.10 \\\hline
  Johann Sebastian Bach & 1.02 \\\hline
  John Williams & 0.93 \\\hline
  Manuel de Falla & 0.80 \\\hline
  Traditional & 0.75 \\\hline
  Osvaldo Golijov & 0.63 \\\hline
  Modest Mussorgsky & 0.63 \\\hline
  Richard Wagner & 0.58 \\\hline
   Leroy Anderson & 0.54 \\\hline
  Wolfgang Amadeus Mozart & 0.53 \\\hline
 \end{tabular}
\end{center}
 \caption{Top 10 composers with the largest posterior mean additive row effects, $a_i$.  The additive row effect is an overall measure of ``popularity'', in terms of the number of other composers performed after these composers.  Many of the composers with the largest row effects (and thus those that were performed frequently) are popular composers.}
 \label{tab:row-effects-large}
\end{table}

\noindent Many of the composers with the smallest posterior mean row random effects are contemporary composers (\autoref{tab:row-effects-small}).  For these composers, the small row random effects indicate that works by many of these composers are rarely performed and reinforces that many new works are unlikely to be performed again after their premiere \citep{Tommasini:2008}.  Additionally, these small row random effects indicate that many contemporary composers are not performed with a variety of other composers.  \newline


\begin{table}
 \begin{center}
 \begin{tabular}{|l|l|l|}
  \hline
 Composer & Year of Birth &   $a_i$  \\
  \hline\hline
  Alexander Glazunov &   1865      & -0.60  \\\hline
  Arnold Schoenberg   &   1874     & -0.59  \\\hline
  Oliver Knussen   &   1952        & -0.54  \\\hline
  Charles Wuorinen & 1938 & -0.50  \\\hline
  Carl Nielsen     &   1865     & -0.49  \\\hline
  Bohuslav Martinu &   1890     & -0.48  \\\hline
  John Harbison  & 1938 & -0.46  \\\hline
  Witold Lutoslawski  & 1913    & -0.43  \\\hline
  Ervin (Erwin) Schulhoff & 1894 & -0.38 \\\hline
  Alfred Schnittke &   1934     & -0.36  \\\hline
 \end{tabular}
\end{center}
 \caption{Top 10 composers with the smallest posterior mean additive row effects, $a_i$.  Many of the composers with the smallest additive row effects are contemporary composers, meaning many recent works are not performed before a large number of other composers.  This reinforces that many new works are not performed again after their premiere.}
 \label{tab:row-effects-small}
\end{table}


\begin{figure}
 \centerline{
 \includegraphics[width=0.8\columnwidth]{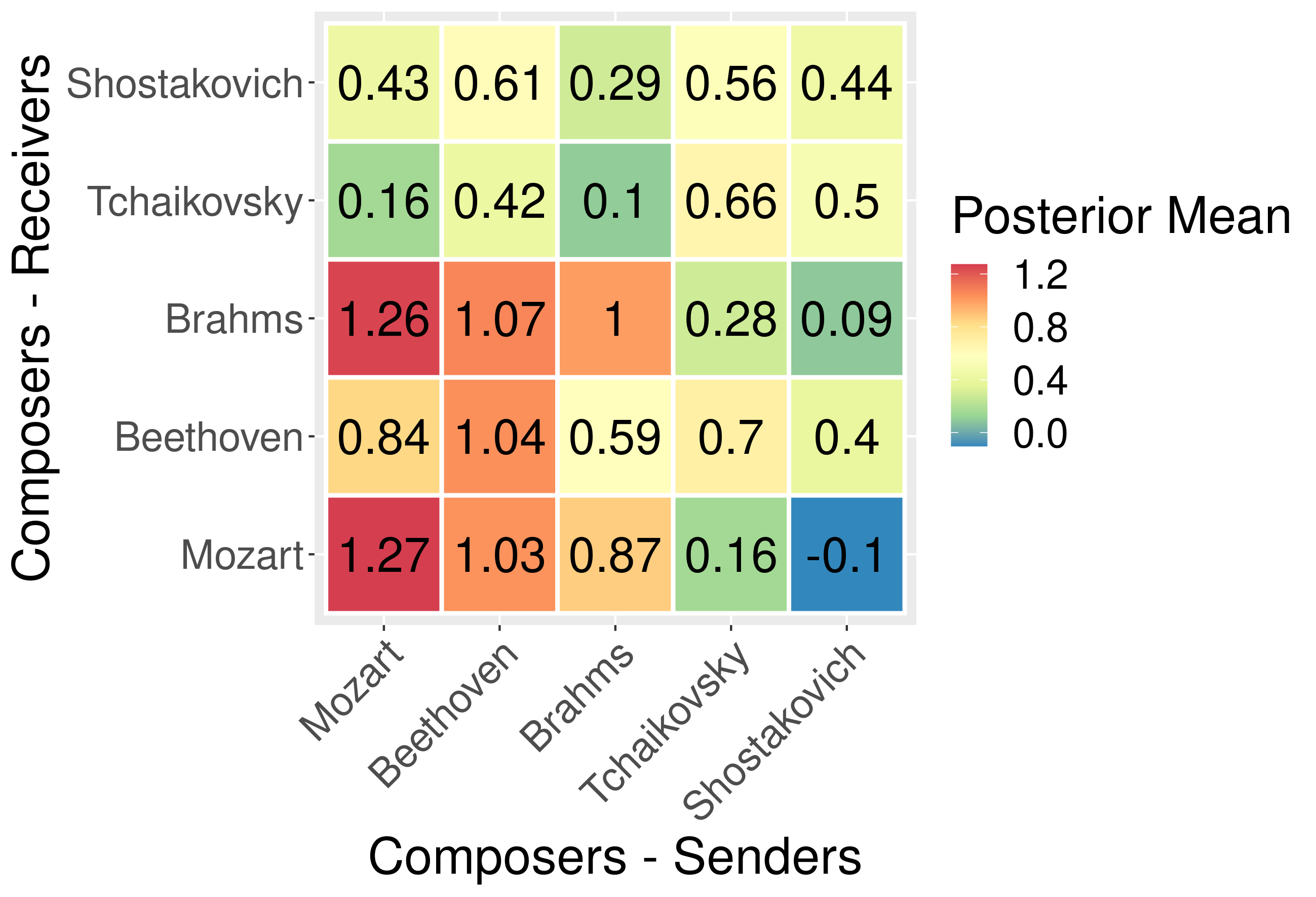}}
 \caption{Heatmap of the multiplicative effects for 5 of the most frequently performed composers.}
 \label{fig:heatmap}
\end{figure}

\noindent Finally, the multiplicative effects can be examined by looking at the posterior mean of the $UV^T$ multiplicative random effects matrix, shown for 5 of the most frequently performed composers in \autoref{fig:heatmap}.  The column for Shostakovich (i.e. the composer sender multiplicative effect for Shostakovich) has low values relative to the other composers considered here.  Works by Shostakovich, especially his symphonies, tend to be longer and rather ``heavy'' or ``dark'' in tone, and were thus less likely to be programmed after a lighter work by Mozart, for example.  Additionally, due to the length, many pieces by Shostakovich were performed at the end of a concert, and thus there were very few composers in general that were performed after Shostakovich.  Shostakovich does have a relatively high in-degree (\autoref{tab:EDA}), but low posterior mean receiver (row) multiplicative random effects for the subset of popular composers in \autoref{fig:heatmap}, indicating that while there were many distinct composers performed \textit{before} Shostakovich, it was unlikely for Shostakovich to be programmed with several of the other popular composers.\newline

\noindent This is in contrast to a composer like Brahms, which has a relatively high in-degree and out-degree (\autoref{tab:EDA}) and higher multiplicative receiver (row) and sender (column) multiplicative effects in \autoref{fig:heatmap}.  Brahms has many shorter works, including frequently performed overtures, that can be programmed before many of the other popular composers, which can be seen in the higher values for the Brahms row in the $UV^T$ matrix in \autoref{fig:heatmap}.  Interestingly, Brahms was unlikely to be programmed before or after Tchaikovsky by the BSO.  Brahms and Tchaikovsky are both famous Romantic era composers and have several pieces that could fit well on the same program from a musical perspective, but based on the low posterior mean for the multiplicative effect matrix, it appears that these two composers are very rarely programmed together by the BSO in practice.\newline

\noindent We can additionally cluster the multiplicative effects $U$ using k-means clustering to explore which composers are similar, based on the model results.  The resulting five clusters make sense from a musical perspective, reflecting that the multiplicative effects in the AME model are able to capture musically logical relationships between composers.  The first cluster contains 62 composers and roughly corresponds to Russian and French Romantic and Modern composers; 1/3 of the composers in this cluster are Russian or French.  A subset of composers in this cluster include: Camille Saint-Saëns, Dmitri Shostakovich, Hector Berlioz, Mikhail Glinka, Modest Mussorgsky, Nikolai Rimsky-Korsakov, Olivier Messiaen, Pyotr Ilyich Tchaikovsky, Richard Strauss and Sergei Prokofiev.  The second cluster roughly corresponds to American Modern composers and Baroque composers, including Aaron Copland, George Gershwin, J. S. Bach, Georg Philipp Telemann, George Frideric Handel,  John Cage, Samuel Barber and Wynton Marsalis.   American composers make up 35\% of this cluster and 3 of the 5 Baroque composers played by the BSO are also in this cluster.\newline  

\noindent Famous, frequently performed composers make up the third cluster, including Antonin Dvorák,  Arnold Schoenberg,  Béla Bartók,  Claude Debussy, Felix Mendelssohn, Franz Joseph Haydn, Franz Schubert, Frédéric Chopin, Gustav Mahler, Igor Stravinsky, Jean Sibelius, Johannes Brahms, Leonard Bernstein, Ludwig van Beethoven, Maurice Ravel, Richard Wagner, Robert Schumann, Sergei Rachmaninoff and Wolfgang Amadeus Mozart.  Finally, the fourth and fifth clusters both approximately represent composers that are frequently performed during Pops concerts; the fourth cluster contains 33 composers, including Giacomo Puccini, Gioachino Rossini, Giuseppe Verdi and Richard Rodgers, while the fifth cluster includes 69 composers, including Alan Menken, Alexander Borodin, Antonio Vivaldi, Georges Bizet and Johann Strauss, Jr. \newline

\noindent The multiplicative effect matrix in \autoref{fig:heatmap} gives interesting insight into how the most popular composers are programmed together.  Clearly, from the perspective of length of piece alone, not all of these popular composers could be programmed together.  A program with a Mahler and Shostakovich symphony is unlikely due to the length of typical symphonies by both of these composers, however, it seems that some of these popular composers could be programmed together in creative ways.  This would allow for the programming of familiar composers and popular works, while creating concerts that were surprising to the audience by the novel combination of composers in the same concert. Analyzing concert programming trends from a model-based perspective can thus provide interesting insights into current trends and suggest areas for further exploration in future programming by orchestras.

\section{Conclusions and Future Work}\label{sec:conclusion}

Overall, AME network models were able to capture relationships between composers in concerts programmed by the BSO over the past 20 years.  Analysis of the model results indicated that the type of piece was the most important in determining if two composers would be programmed together, while the era and the nationality of the composer were much less important.  Additionally, the results emphasized that works by many contemporary composers are unlikely to be performed again after their premiere.  Finally, the multiplicative effects in the AME model allowed for direct analysis of the relationships between composers in terms of programming, and found that not all popular composers are programmed together with the same likelihood.\newline

\noindent There are several directions for future work.  First, it would be interesting to add additional covariates to the analysis.  From a musical perspective, covariates such as the length of each piece, instrumentation, whether there is a soloist or not and more subjective descriptions of each composer, such as musical tone, would be helpful to further explore which factors are most important in concert programming. This would also allow for a closer examination of how important the type of piece is, when the length of the piece, for example, is directly accounted for.  Additionally, considering dynamic network models would allow for exploring how programming practices have changed over time. Comparisons of trends across different orchestras, or in the context of a hierarchical model, could also be explored, as several major orchestras have detailed online archives.  Of most interest, however, would be to include revenue or attendance data to explore how programming influences audience engagement and potentially even provide recommendations for future concert programs.  However, this initial analysis shows the promise in using statistical network models to analyze orchestral concert programming and the ability of the models to recover well-known musical trends for concert programming.

\break\newpage
\bibliographystyle{plainnat}
\bibliography{ISMIR}

\end{document}